\documentclass[preprintnumbers, prd, showpacs, onecolumn, floatfix, preprintnumbers, letterpaper, amsmath, amssymb, superscriptaddress]{revtex4}

\usepackage{amssymb, amsmath, bm, dcolumn, epsf, graphicx, latexsym, mathbbol, slashed}

\usepackage{color}
\newcommand{\tc}{\textcolor{black}}

\newcommand{\be}{\begin{equation}}
\newcommand{\ee}{\end{equation}}
\newcommand{\bea}{\begin{eqnarray}}
\newcommand{\eea}{\end{eqnarray}}
\newcommand{\barr}{\begin{array}}
\newcommand{\earr}{\end{array}}

\begin{document}
\hfill USTC-ICTS-14-07
\title{Fluctuations of cosmological birefringence and the effect on CMB B-mode polarization}

\author{Wen Zhao}
\affiliation{Key Laboratory for Researches in Galaxies and Cosmology, Department of Astronomy, University of Science and Technology of China, Hefei, Anhui,
230026, China}

\author{Mingzhe Li}
\affiliation{Interdisciplinary Center for Theoretical Study, University of Science and Technology of China, Hefei, Anhui 230026, China}


\begin{abstract}
{The cosmological birefringence caused by the coupling between the
cosmic scalar field and the cosmic microwave background (CMB)
photons through the Chern-Simons term can rotate the polarization
planes of the photons, and mix the CMB E-mode and B-mode
polarizations. The rotation angle induced by the dynamical scalar field can be separated into the
isotropic background part and the anisotropic fluctuations. The effect of the
background part has been be studied in the previous work (Zhao \&
Li, arXiv:1402.4324). In this paper, we focus on the influence of
the anisotropies of the rotation angle. {\tc{ We first assume that the cosmic scalar field is massless,
consistent with other works, we find that the rotation
spectrum can be quite large, which may be detected by the
potential CMB observations. However, if the scalar field is identified as the quintessence field,
by detailed discussing both the entropy and adiabatic perturbation modes for the first time,
we find that the anisotropies of the rotation angle are always too small to be
detectable.}} {\tc {In addition, as the main goal of this paper,
we investigate the effect of rotated polarization power spectrum on the detection of relic gravitational waves.
we find that, the rotated
B-mode polarization could be fairly large, and comparable with
those generated by the gravitational waves. This forms a new contamination for the detection of
relic gravitational waves in the CMB. In particular, we also
propose the method to reconstruct and subtract the rotated B-mode
polarization, by which the residuals become negligible for the
gravitational-wave detection.}}

}
\end{abstract}


\maketitle

\section{Introduction}

Inflation is the most popular scenario of the extremely early Universe \cite{guth1981}. In addition to elegantly solve the flatness puzzle, the horizons puzzle and the monopoles puzzle in the hot bag-bang universe, inflationary models predict the primordial scalar and tensor fluctuations with the nearly scale-invariant power spectra \cite{inflation-perturbation}. The scalar perturbations seeded the large-scale structure, which is highly consistent with the current observations on temperature and polarization anisotropies of the cosmic microwave background (CMB) radiation \cite{wmap,planck0} and the distributions of the galaxies in various large-scale structure observations. The primordial tensor perturbations, i.e., relic gravitational waves, encoded all the evolution information of the Universe \cite{grishchuk}. The amplitude corresponds to the energy scale of inflation, and the spectral index reflects the evolution of scale factor in the inflation stage. So, the detection of relic gravitational waves is always treated as the smoking-gun evidence of the inflation, which mainly depends on the observation of the CMB polarization, in particular the so-called B-mode polarization \cite{ref-Bmode}. The current observations, including those of WMAP and Planck missions, are yet to detect a definite signal of relic gravitational waves. However, the recent observations of BICEP1, SPTPOL and POLARBEAR telescopes have given some interesting results for the B-mode polarizations \cite{bicep,spt,polarbear}, which encourage us to put the gravitational-wave detection through the CMB polarizations as a highest priority task for the next generations of CMB experiments \cite{task}.

The CMB B-mode polarization is contaminated by many sources, including the instrumental noises, various foreground emissions, the E-B mixture due to the partial sky observations, the leakage of E-mode polarizations caused by the cosmic weak lensing, as well as the cosmic birefringence. It is well known that the cosmic birefringence can be caused by the possible interaction between the photons and the cosmic scalar field $\varphi$ in the Universe \cite{carroll1990,kamionkowski1998}. The birefringence induces rotation of the polarization plane of CMB, and converts E-mode and B-mode polarization, which forms a new contamination for the gravitational-wave detection. Recently, numerous attentions have been attracted to study this effect from both theoretical and observational sides \cite{other-cpt,other-constraints}.

The cosmic scalar field may or may not be identified as the quintessence dark energy. {\tc{ In the previous works \cite{li2008,kamionkowski2008,kamionkowski2009,zaldarriaga2009,Caldwell:2011}, the main attentions have been payed to the case, where the scalar field is massless. Consistent with these works, we find that, although the background rotation angle is absent in this case, the spatial fluctuations could be quite large, and could be well detected by the potential observations of Planck satellite, CMBPol mission or some other experiments. In this paper, we also consider the case where the scalar field is identified with the quintessence dark energy, and the background rotation angle can be large. For the first time, we detailed discuss both the entropy and the adiabatic perturbation modes, and find that the fluctuations in this case are always very small, and beyond the future detection abilities.}}

{\tc{As the main task of this paper, we focus on the influence of the cosmological birefringence on the detection of relic gravitational waves. For the former case, where the cosmic scalar field is massless, we investigate in detail the rotated B-mode polarization, and find they can be fairly large and comparable with (or even larger than) those caused by relic gravitational waves or cosmic weak lensing. So, it is important to reduce it in the future gravitational-wave detections. By utilizing the statistical properties of the reconstructed rotation angle coefficients $\alpha_{\ell m}$ and those of the E-mode coefficients $E_{\ell m}$, we propose the method to reconstruct the rotated B-mode polarization. We find that the rotated polarization can be reconstructed and removed at the very high level, if considering the noise levels of CMBPol mission or the better experiments, which become completely negligible for the detection of relic gravitational waves in the CMB.}}

The outline of this paper is as follows.
In Sec. \ref{sec2}, we briefly introduce the cosmological birefringence and focus on the rotation angle fluctuations.
In Sec. \ref{sec3}, we discuss the fluctuations of the rotation angle in both the massless scalar field and the quintessence field models.
In Sec. \ref{sec4}, we investigate the possible detection of the fluctuations of the rotation angle by the potential CMB observations.
In Sec. \ref{sec5}, we discuss the reconstruction and subtraction of the rotated B-mode polarization, and their effect on the gravitational-wave detection.
Sec. \ref{sec6} summarizes the main results of this paper.

\section{Anisotropies of rotation angle \label{sec2}}


The cosmological birefringence can be caused by the coupling of CMB photons to a cosmic scalar field $\varphi$ through the Chern-Simons term. Let us consider the theory with the following Lagrangian \cite{li2008,Caldwell:2011},
 \begin{eqnarray}\label{csL}
 \mathcal{L}=-\frac{1}{2}(\partial_{\mu}\varphi)(\partial^{\mu}\varphi)-V(\varphi)-\frac{1}{4}F^{\mu\nu}F_{\mu\nu}-\frac{\beta\varphi}{2M}F_{\mu\nu}\tilde{F}^{\mu\nu},
 \end{eqnarray}
where $\tilde{F}^{\mu\nu}=\epsilon^{\mu\nu\rho\sigma}F_{\rho\sigma}/2$ is the dual of the electromagnetic tensor, $\epsilon^{\mu\nu\rho\sigma}$ is the
Levi-Civita tensor, $M$ is a mass scale for the effective field theory, and $\beta$ is the dimensionless coupling. The last Chern-Simons term violates the CPT symmetry if the scalar field $\varphi$ develops a background. It introduces a modification of Maxwell's equations that results in different dispersion relations for left and right-circularly polarized photons. Consequently,
linearly-polarized electromagnetic waves that propagate over cosmological distances undergo cosmological birefringence, a frequency-independent rotation of
the plane of polarization by an angle $\alpha$. The value of rotation angle is calculated by \cite{carroll1990,li2008}
 \begin{eqnarray}\label{alpha}
 \alpha=\frac{\beta}{M}\Delta\varphi,
 \end{eqnarray}
where $\Delta\varphi$ is the change in $\varphi$ over the photon's trajectory. For the CMB, the polarization rotation is determined by the change in
$\varphi$ since the decoupling time.

Since $\varphi$ is a dynamical field,  it must have spatial
fluctuations. This leads to a direction-dependent rotation angle
in (\ref{alpha}). As we have usually done in the cosmological
perturbation theory, the scalar field can be separated into the
homogeneous background part and the inhomogeneous fluctuations
which are treated to be small in the linear perturbation theory.
Hence the rotation angle led by the scalar field can be also
separated into the isotropic background part and the anisotropic
fluctuations, which are considered to be randomly distributed on
the sky  \cite{li2008}:
 \begin{eqnarray}\label{full_alpha}
 \alpha=\bar{\alpha}+\delta\alpha,
 \end{eqnarray}
where
 \begin{eqnarray}\label{alpha0}
 \bar{\alpha}=\frac{\beta}{M}(\bar{\varphi}(\tau_0)-\bar{\varphi}(\tau_{dec})),
 \end{eqnarray}
 \begin{eqnarray}\label{alpha1}
 \delta\alpha(\hat{\bf n})=-\frac{\beta}{M}\delta\varphi(\vec{x}_{dec}, \tau_{dec}).
 \end{eqnarray}
In the above equations, $\tau_0$ denotes the present conformal time, and $\tau_{dec}$ is the conformal time at the matter-radiation decoupling. The effect of the isotropic rotation angle $\bar{\alpha}$ have been detailedly discussed by many authors \cite{other-cpt,zhao2014}. In this paper, we only focus on the anisotropic rotation $\delta\alpha(\hat{\bf n})$, which is determined by Eq.(\ref{alpha1}).  For this study it is very convenient to use the synchronous gauge, $\delta\varphi\equiv(\delta\varphi)_{\rm syn}$. As pointed out in \cite{Caldwell:2011}, the CMB photons from all directions experienced the last scatterings at some fixed temperature, and the perturbation $\delta\varphi$ on the surface of this constant CMB temperature determines the anisotropies of the rotation angle. In the synchronous gauge, the surface of constant temperature is identical with that of constant time. So, except for the Appendix, in the main text of this paper, we shall work in the synchronous gauge.

The rotation angle $\delta \alpha$ is treated as a two-dimensional spherical fluctuation field. We expand it via the spherical harmonics as follows:
 \begin{eqnarray}\label{expand_delta_alpha}
 \delta\alpha (\hat{\bf n})=\sum_{\ell m} \alpha_{\ell m}Y_{\ell m}(\hat{\bf n}),
 \end{eqnarray}
where $\alpha_{\ell m}$ are the coefficients of decomposition. If the perturbation field satisfies the Gaussian distribution, the statistical properties are entirely described by the angular power spectrum, which is defined as
 \begin{eqnarray}\label{alpha_power}
 \langle \alpha^*_{\ell m} \alpha_{\ell'm'}\rangle=C^{\alpha\alpha}_{\ell}\delta_{\ell\ell'}\delta_{mm'},
 \end{eqnarray}
where we have assumed statistical isotropy of $\alpha_{\ell m}$. From Eq.(\ref{alpha1}), we find the angular power spectrum $C_{\ell}^{\alpha\alpha}$ for
the rotation angle is calculated by \cite{li2008}
 \begin{eqnarray}\label{clalpha}
 C_{\ell}^{\alpha\alpha}=4\pi \left(\frac{\beta}{M}\right)^2 \int \frac{dk}{k} P_{\delta\varphi}(k) j^2_{\ell}(k\Delta\tau),
 \end{eqnarray}
where $j_l(x)$ is the spherical bessel function and $\Delta \tau=\tau_0-\tau_{dec}$, $P_{\delta\varphi}(k)$ is the power spectrum of perturbation $\delta\varphi$ at the decoupling time, which is defined by $\langle \delta\varphi_k^*(\tau_{dec})\delta\varphi_k(\tau_{dec})\rangle=\frac{2\pi^2}{k^3}P_{\delta\varphi}(k)$. In the next section, we shall calculate the spectrum $P_{\delta\varphi}(k)$ in various models.


\section{Evolution of scalar field and power spectrum of anisotropic rotation \label{sec3}}

We consider the spatially flat universe. In the synchronous gauge, the metric perturbation is given by the following line element \cite{Ma:1995}
 \begin{eqnarray}\label{newtonian_gauge}
 ds^2=a^2(\tau)\{-d\tau^2+(\delta_{ij}+h_{ij})dx^idx^j\}.
 \end{eqnarray}
The scalar metric perturbations $h_{ij}$ can be decomposed into a trace part $h\equiv h_{ii}$ and traceless part $h_{ij}^{||}$, i.e. $h_{ij}=h\delta_{ij}/3+h_{ij}^{||}$. By definition, the divergence of $h_{ij}^{||}$ is longitudinal, which follows that $h_{ij}^{||}$ can be written in terms of a scalar field $\mu$ as follows:
 \begin{eqnarray}
 h_{ij}^{||}=\left(\partial_{i}\partial_{j}-\frac{1}{3}\delta_{ij}\nabla^2\right)\mu.
 \end{eqnarray}
So, the two scalar fields $h$ and $\mu$ characterize the scalar modes of the metric perturbations. In general, we shall work in the Fourier $k$-space. We introduce two fields $h_{\vec{k}}(\tau)$ and $\eta_{\vec{k}}(\tau)$ in $k$-space and write the scalar mode of $h_{ij}$ as a Fourier integral
 \begin{eqnarray}
 h_{ij}(\vec{x},\tau)=\int d^3 k e^{i\vec{k}\cdot\vec{x}} \left\{\hat{k}_i \hat{k}_j h_{\vec{k}}(\tau)+(\hat{k}_i \hat{k}_j-\frac{1}{3}\delta_{ij})6\eta_{\vec{k}}(\tau)\right\}.
 \end{eqnarray}
Since in this paper, we assume the statistical isotropy of the perturbations, so in the following discussions, $h_{\vec{k}}$ and $\eta_{\vec{k}}$ can be simplified as $h_{{k}}$ and $\eta_{{k}}$, respectively.


Let us consider the scalar field $\varphi$ in the Universe, which can be separated as the background part $\bar{\varphi}$ and the first-order perturbation $\delta \varphi$, i.e. $\varphi=\bar{\varphi}+\delta\varphi$. From the action (\ref{csL}) we have the equation of motion for $\varphi$,
\be
\Box\varphi+\frac{dV}{d\varphi}=0~,
\ee
where we have neglected the small back reaction of photons from the Chern-Simons coupling.
The equations for the background part and the perturbations are obtained from the above Klein-Gordon equation directly,
 \begin{eqnarray}\label{varphi0}
 \bar{\varphi}''+2\mathcal{H}\bar{\varphi}'+a^2V_{\bar{\varphi}}=0,
 \end{eqnarray}
 \begin{eqnarray}\label{varphi1}
 \delta\varphi''&+&2\mathcal{H}\delta\varphi'-\nabla^2\delta\varphi+a^2V_{\bar{\varphi}\bar{\varphi}}\delta\varphi =-\frac{1}{2}h'\bar{\varphi}',
 \end{eqnarray}
where $prime$ denotes $d/d\tau$, $\mathcal{H}=a'/a$, $V_{\bar{\varphi}}\equiv dV/d\bar{\varphi}$ and $V_{\bar{\varphi}\bar{\varphi}}\equiv d^2V/d\bar{\varphi}^2$. In principle, the value of $\delta\varphi(\tau_{dec})$ can be solved by using this equation, which depends on the evolution of the cosmic scalar factor $a(\tau)$, the perturbation $h(\tau)$, the potential $V(\varphi)$, and the background part of scalar field $\bar{\varphi}$.  In this paper, we shall consider two cases for the cosmic scalar field. One is the massless scalar field, and the other is the case where the scalar field acts as the cosmic quintessence, which drives the acceleration of the current Universe.

\subsection{Massless scalar field}

\begin{figure}
\begin{center}
\includegraphics[width=10cm]{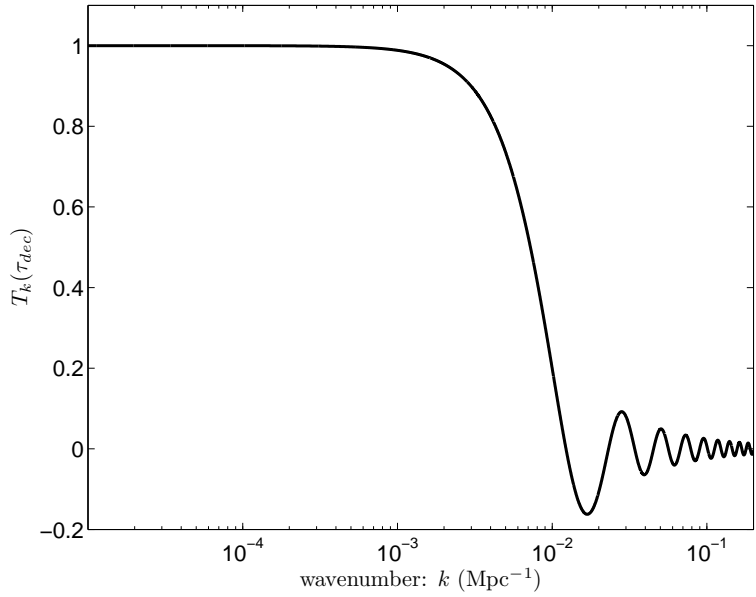}
\caption{\label{figb0} The transfer function $T_k(\tau_{dec})$ as a function of wavenumber $k$.}
\end{center}
\end{figure}

Similar to the previous works
\cite{pospelov2008,zaldarriaga2009,Caldwell:2011}, in the first
scenario, we consider the simple case where the scalar field
$\varphi$ is supposed to be a massless scalar field with
$V(\varphi) \equiv 0$. In this case, the evolution equations in
Eqs.(\ref{varphi0}) and (\ref{varphi1}) are simplified as follows,
 \begin{eqnarray}\label{varphi0a}
 \bar{\varphi}_k''+2\mathcal{H}\bar{\varphi}_k'=0,
 \end{eqnarray}
 \begin{eqnarray}\label{varphi1a}
 \delta\varphi_k''&+&2\mathcal{H}\delta\varphi_k'+k^2\delta\varphi_k
 =-\frac{1}{2}h_k'\bar{\varphi}_k',
 \end{eqnarray}
where the evolution equations have been written in the Fourier
space. Obviously, the solution of Eq.(\ref{varphi0a}) is
$\bar{\varphi}'_k=0$, which predicts no uniform rotation
$\bar{\alpha}$. Inserting $\bar{\varphi}_k'=0$ into
Eq.(\ref{varphi1a}), we find that
 \begin{eqnarray}\label{varphi1aa}
 \delta\varphi_k''&+&2\mathcal{H}\delta\varphi_k'+k^2\delta\varphi_k=0,
 \end{eqnarray}
the perturbation $\delta\varphi$ is decoupled from the metric
perturbations and always corresponds to the entropy perturbation.
The solution of this equation can be formally written as
$\delta\varphi_k(\tau)=\delta\varphi_k(\tau_{ini})T_k(\tau)$,
where $T_k(\tau)$ is the transfer function. The initial condition
can be set at the beginning of the radiation-dominant stage, where
$\delta\varphi'_k(\tau_{ini})=0$ and
$P_{\delta\varphi}(k,\tau_{ini})=H_I^2/(4\pi^2)$.
$H_I$ is the Hubble parameter in the inflationary stage. Instead
of using the analytic approximation
\cite{zaldarriaga2009,Caldwell:2011}, here we numerically solve
the transfer function, and present $T_k(\tau_{dec})$ as a function
of $k$ in Fig.{\ref{figb0}}. When $k\tau_{dec} \ll 1$, we have
$T_k(\tau_{dec})=1$. While $k\tau_{dec}\gg 1$, the transfer
function begins to oscillate, and the amplitude is damping with
the expansion of the Universe.

\begin{figure}
\begin{center}
\includegraphics[width=10cm]{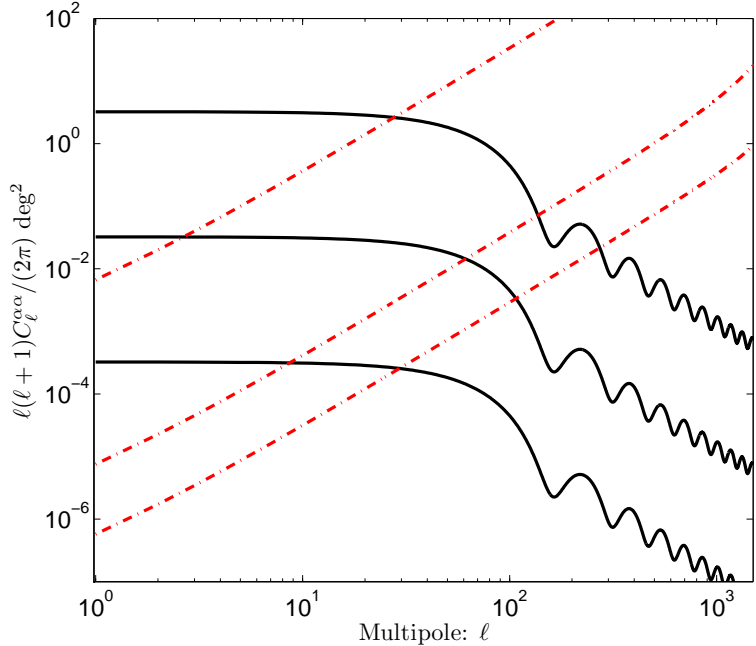}
\caption{\label{figdd1} The solid lines denote the rotation-angle power spectrum of the massless scalar field case with $\beta H_I/M=0.2$ (upper black line), $0.02$ (middle blue line), $0.002$ (lower green line), respectively. The red dashed lines denotes the corresponding noise power spectrum $\ell(\ell+1){N}_{\ell}^{\alpha\alpha}/(2\pi)$. The upper one is for the Planck noise case, the middle one is for the CMBPol noise case, and the lower one is for the case of the reference experiment.}
\end{center}
\end{figure}

Inserting $\delta\varphi_k(\tau_{dec})$ into Eq.(\ref{clalpha}), we can obtain the power spectrum of the rotation angle $C_{\ell}^{\alpha\alpha}$. In the previous works \cite{pospelov2008,12065546,li2013}, the authors have considered the constraints on the fluctuation of the rotation angle from the CMB observations. In addition, in \cite{kamionkowski2010}, the authors constrained it by the active galactic nuclei data, and obtain the upper limit of the variance of $\delta\alpha$ as follows $\sqrt{\langle(\delta\alpha)^2\rangle}\lesssim 3.7^{\circ}$. In this paper, we shall consider it as the current observational upper limit. In the massless scalar field case, it follows that combined constraint on the cosmological parameters $\beta$, $M$ and $H_I$,
 \begin{equation}
 \beta H_I/M \lesssim 0.2.
 \end{equation}
In Fig. \ref{figdd1}, we plot the power spectrum
$C_{\ell}^{\alpha\alpha}$ for the cases with $\beta H_I/M=0.2$,
$0.02$, $0.002$, respectively. We find that the values of
$C_{\ell}^{\alpha\alpha}$ rapidly decrease for the high
multipoles. So we anticipate the potential detection will mainly
be in the low multipole $\ell<100$. In this range, the rotation
power spectrum can be written as,
 \begin{equation}
 \ell(\ell+1)C_{\ell}^{\alpha\alpha}=40.56\left(\frac{\beta H_I/M}{0.2}\right)^2~{\rm deg}^2.
 \end{equation}

\subsection{Quintessence}

Now, let us consider the case, in which the scalar field is identified as the cosmological quintessence field. Different from the massless scalar field case, the average rotation angle in this case can be fairly large, which has been realized for a long time \cite{kamionkowski1998}. Here, we shall discuss in detail the anisotropies of the rotation angle. The evolution of $\delta\varphi$ is governed by Eq. (\ref{varphi1}). Because what we will solve is the evolution of $\delta\varphi$ from the time well into the radiation dominated era to the recombination which takes place at matter dominated epoch, the quintessence was subdominant during this period and its contributions to the background and the metric perturbations are negligibly small. Analytically we may think the metric perturbations are fixed by the dominant fluids and the back reaction from the quintessence can be neglected. {\tc{The solution to the full equation (\ref{varphi1}) has two origins, one is the general solution to the homogeneous equation and another one is the special solution to the inhomogeneous equation. The former belongs to a two-parameter family specified by the initial conditions $\delta\varphi_k$ and $\delta\varphi'_k$, and are unaffected by the perturbations of the metric and other species, which corresponds to the entropy perturbation or entropy mode. The latter is sourced by the metric perturbation and after enough long time it will be unrelated with the initial values of $\delta\varphi_k$ and $\delta\varphi'_k$, which is the adiabatic mode. This is similar to the forced oscillator, which can oscillate with its intrinsic frequency only relying on the parameters of the oscillator itself and with the same frequency with the external driving force if it is periodic. Generally the motion of the forced oscillator is the superposition of these two modes. The solution to Eq. (\ref{varphi1}) is the superposition of the entropy mode and the adiabatic mode. In the previous works \cite{Caldwell:2011,lee}, the authors have focused only on the adiabatic mode of several special quintessence models. Different from them, in this paper, we shall analytically calculate and discuss the general properties of both perturbations modes for the general quintessence models, and investigate the rotation angle power spectrum and the detection possibilities. }}


\subsubsection{Solution to the homogeneous equation: entropy mode}

The homogeneous equation for $\delta\varphi$
 \begin{eqnarray}\label{varphi11}
 \delta\varphi_k''+2\mathcal{H}\delta\varphi_k'+(k^2+a^2V_{\bar{\varphi}\bar{\varphi}})\delta\varphi_k = 0,
 \end{eqnarray}
is slightly different from the equation (\ref{varphi1aa}) for the massless field due to the existence of the new term $a^2V_{\bar{\varphi}\bar{\varphi}}$ from the potential. For the case in which $k^2\gg a^2V_{\bar{\varphi}\bar{\varphi}}$ is satisfied throughout the evolution, this evolution equation returns to that of massless scalar field in Eq.(\ref{varphi1aa}). The typical example of this case is the slow-rolling quintessence models, where the kinematic energy is much smaller than the potential energy, and the equation-of-state of quintessence is very close to $-1$. As in general,the slow-rolling parameter $\eta$ can be defined as follows
 \begin{equation}
 \eta \equiv \frac{1}{8\pi G}\frac{V_{\bar{\varphi}\bar{\varphi}}}{V},
 \end{equation}
which is much smaller than $1$ in the slow-rolling stage. So we
have $a^2 V_{\bar{\varphi}\bar{\varphi}}=4\pi
Ga^2(1-w)\rho_{\varphi}\eta$. For the slow-rolling quintessence
field, $w\simeq -1$ and $\rho_{\varphi}$ nearly keeps constant, so
we have
 \begin{equation}
 a^2 V_{\bar{\varphi}\bar{\varphi}} \simeq {3}a^2\eta\Omega_{\varphi}H_0^2\ll H_0^2,
 \end{equation}
where $\Omega_{\varphi}$ is the current energy density parameter
of quintessence component, and $H_0$ is the present Hubble
parameter, $a$ is the scale factor of the Universe, which is
normalized as $a_0=1$. We should remember that, for the frequency
modes of perturbation $\delta\varphi$, which can lead to the
observable cosmological birefringence, should satisfy $k^2 \ge
H_0^2$, i.e. reentered the horizon in the present stage. So we
conclude that for the slow-rolling quintessence field, we have
$k^2\gg a^2V_{\bar{\varphi}\bar{\varphi}}$, and the homogeneous
solution of the evolution equation returns to the case of massless
scalar field.

Now, let us evaluate the amplitude the power spectrum for the entropy perturbation. The initial conditions are assumed to be the same with the massless field, i.e., $\delta\varphi_k'(\tau_{ini})=0$ and $P_{\delta\varphi}(k,\tau_{ini})=H_I^2/(4\pi^2)$. This is reasonable if the quintessence has been existed during inflation and was effectively massless. From the previous subsection, we know that the main power spectra are at the low multipoles $\ell \lesssim 100$. And in this range, we have $\ell(\ell+1)C_{\ell}^{\alpha\alpha}=40.56(\frac{\beta H_I/M}{0.2})^2~{\rm deg}^2$. The inflationary Hubble parameter $H_I$ relates the tensor-to-scalar ratio by $H^2_I=1.2\times 10^{-8}r$ (Note that, in this paper, we have set the reduced Planck mass $M_{\rm pl}=1$.). The tightest constraint $r<0.11$ \cite{planck}, obtained from the Planck observations, follows that $H_I^2 \lesssim 10^{-9}$. On the other hand, in the quintessence models, the uniform rotation $\bar{\alpha}$ is also predicted as we have mentioned, and its values is determined by Eq.(\ref{alpha0}), where the value of $(\bar{\varphi}(\tau_0)-\bar{\varphi}(\tau_{dec}))\simeq \bar{\varphi}(\tau_0)$ is always close to the Planck mass, and the current constraint of $\bar{\alpha}$ is $|\bar{\alpha}|\lesssim O(1^{\circ})$ \cite{other-constraints}. So we have $(\beta/M)^2 \lesssim O(10^{-3})$. Combining all these results, we find that the amplitude of the power spectrum in the low multipoles are $\ell(\ell+1)C_{\ell}^{\alpha\alpha} \lesssim 10^{-10}~{\rm deg}^2$, which is too small to be detectable.

Now, let us turn to the general quintessence models. We can express the second derivations of the potential in terms of the equation-of-state of quintessence $w$ as follows \cite{dave2002}
 \begin{eqnarray}\label{v-aa}
 a^2 V_{\bar{\varphi}\bar{\varphi}}=-\frac{3}{2}(1-w)\left[\frac{a''}{a}-\mathcal{H}^2\left(\frac{7}{2}+\frac{3}{2}w\right)\right]
                       +\frac{1}{1+w}\left[\frac{w'^2}{4(1+w)}-\frac{w''}{2}+w'\mathcal{H}(3w+2)\right].
 \end{eqnarray}
We assume the time evolution of $w$ is small, and $w\nrightarrow -1$ (i.e. $1+w=O(1)$) before the decoupling stage. For these models, in both radiation-dominant and matter-dominant stages, we have that
 \begin{eqnarray}
 a^2 V_{\bar{\varphi}\bar{\varphi}} \simeq \frac{3\mathcal{H}^2}{4}(1-w)\left({7}+{3}w\right).
 \end{eqnarray}
And its value is proportional to $\tau^{-2}$. So for any given wavenumber $k$, we always have $a^2 V_{\varphi\varphi}>k^2$ in some early stage. Thus, the homogeneous equation can be simplified as
 \begin{eqnarray}\label{varphi11}
 \delta\varphi_k''&+&2\mathcal{H}\delta\varphi_k'+a^2V_{\varphi\varphi}\delta\varphi_k = 0.
 \end{eqnarray}
In the radiation-dominant stage, the solution is $\delta\varphi_k\propto \tau^{\frac{1}{2}(-1\pm\sqrt{1-4d})}$, where $d\equiv\frac{3}{2}(1-w)(7+3w)$.
Similarly, in the matter-dominant stage, the solution is $\delta\varphi_k\propto \tau^{\frac{1}{2}(-3\pm\sqrt{9-4d})}$. In both cases, we find the solutions are decaying for any $-1<w<1$, which follow that $\delta\varphi_k(\tau_{dec})\rightarrow 0$, i.e. the homogeneous solutions for these models decayed to zero. These have also been numerically checked for the special cases with constant equation-of-state $w$.

\subsubsection{Solution to the full equation: mix of entropy and adiabatic modes}

In this case, one should solve the inhomogeneous equation in Eq.(\ref{varphi1}). In the $k$-space, it is written as
 \begin{eqnarray}\label{varphi1aaa}
 \delta\varphi_k''&+&2\mathcal{H}\delta\varphi_k'+(k^2+a^2V_{\bar{\varphi}\bar{\varphi}})\delta\varphi_k =-\frac{1}{2}h_k'\bar{\varphi}'.
 \end{eqnarray}
The source term $-\frac{1}{2}h_k'\bar{\varphi}_k'$ plays a crucial role for the evolution of $\delta\varphi_k$, which depends on both the metric perturbation $h_k$ and the background evolution of the quintessence $\bar{\varphi}$. The evolution of $\bar{\varphi}$ is governed by the equation
 \begin{equation}\label{varphi0aaa}
 \bar{\varphi}''+2\mathcal{H}\bar{\varphi}'+a^2V_{\bar{\varphi}}=0.
 \end{equation}
Both equations depends on the dark energy models by the potential function $V(\varphi)$. The adiabatic mode corresponds to the special solution the Eq. (\ref{varphi1aaa}). As showed in Appendix A, in the synchronous gauge the adiabatic mode $\delta\varphi_k$ in the radiation and matter dominated eras nearly vanish at the super-horizon scales as long as the quintessence energy density is extremely small. This can be understood as follows. For the adiabatic perturbation, at large scales any two scalars such as the $\varphi$ and the temperature $T$ of CMB have the relation $\delta \varphi/\varphi'=\delta T/T'$. In the synchronous gauge, as we mentioned previously, the surface  $\tau=const.$ is identical with the surface $\delta T=0$, it is also identical with the surface $\delta\varphi=0$ for the adiabatic perturbation. In the real case the quintessence density is not zero, as we shall do in the numerical computations, there should be a small adiabatic mode. In the previous works \cite{Caldwell:2011,lee}, the authors have discussed two special quintessence models, i.e., the pseudoscalar-Nambu-Goldstone potential in \cite{Caldwell:2011} and the typical exponential potentials and power-law potentials \cite{lee}, and found that under the initial adiabatic conditions (no entropy perturbations $\delta\varphi_k(\tau_{ini})=0$) the adiabatic mode in all these models are very small, and beyond the detection abilities of the potential observations.
Different from them, we shall consider nonzero initial perturbation and attributed its origin from inflation as we assumed in the previous subsection. The solution includes both the entropy mode and the small adiabatic mode.  We will consider a large class of quintessence models, whose equation-of-state can be parameterized as the standard Chevallier-Polarski-Linder form \cite{cpl},
 \begin{equation}
 w(a)=w_0+w_a(1-a),
 \end{equation}
where $w_0$ is the present equation-of-state of quintessence, which is close to $-1$ from the current observations. In our calculations, we choose $w_0=-0.99$. $w_a$ describes the evolution of $w$. In the early Universe, especially in the stage before the CMB decoupling, we have $a\rightarrow 0$, and $w=w_0+w_a$. Note that, this parameterized form can approximate well a large of dark energy models, including the tracking and the thawing quintessence models, and has been widely used in the dark energy studies. For the parameterized quintessence models, the second derivations of the potential can be written as in Eq.(\ref{v-aa}), and the first derivations is written as \cite{dave2002}
 \begin{equation}\label{v-a}
 a^2 V_{\bar{\varphi}}=-\frac{1}{2}\left(3\mathcal{H}(1-w)+\frac{w'}{1+w}\right)\bar{\varphi}'.
 \end{equation}

\begin{figure}
\begin{center}
\includegraphics[width=10cm]{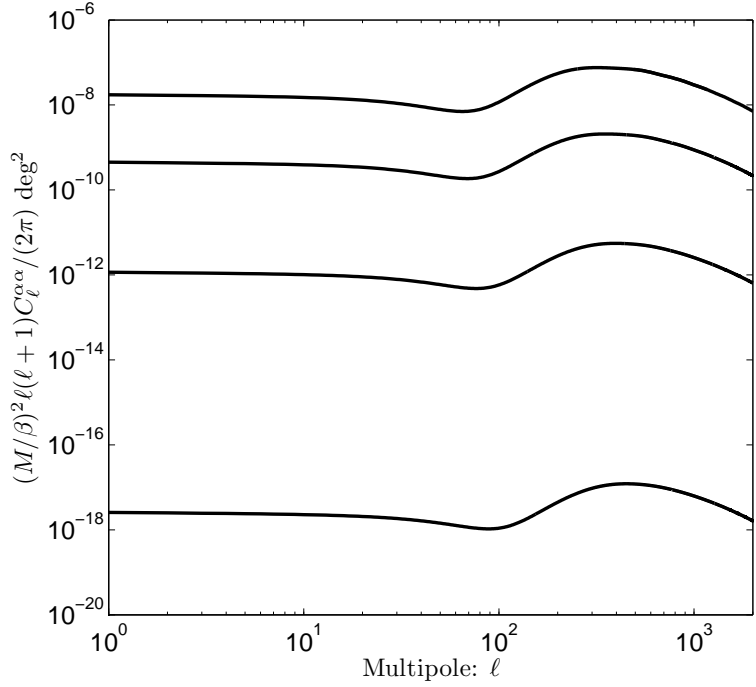}
\caption{\label{figbbbb1} The rotation-angle power spectrum of the quintessence models.
In all the cases, we have assumed the parameterized dark energy model with $w=w_0+w_a(1-a)$ and $w_0=-0.99$. The curves from bottom to top, we have considered $w_a=0.0$, $0.49$, $0.79$, $0.99$, respectively.}
\end{center}
\end{figure}

Considering all these relations, in Appendix A, we numerically solve the equation of $\delta\varphi_k$, and obtain the power spectrum of the rotation angle, which are presented in Fig.\ref{figbbbb1}. In this figure, we have shown the cases with $w_a=0$, $0.49$, $0.79$, $0.99$. In the early Universe, these correspond to the equation-of-state of quintessence $w=-0.99$, $-0.5$, $-0.2$, $0$, respectively. The first case can be identified as a typical slow-rolling dark energy model. While, the last case is similar to the tacker field, which tracked the dark matter component in the early Universe \cite{track}. Note that, if $w_a>0.99$, i.e. $w>0$ in the early stage, the quintessence component would dominate all the other components in some early stage, which may be conflicted with the observations. So we do not consider them in this paper.

From this figure, we find that, in all these cases, the power spectra $C_{\ell}^{\alpha\alpha}$ are too small to be detectable. In particular, for the slow-rolling case with $w_a=0$, we have that $(M/\beta)^2\ell(\ell+1)C_{\ell}^{\alpha\alpha}/2\pi \ll 10^{-16} {\rm deg}^2$.  Even if in the optimal case with $w_a=0.99$, i.e., the  dark-matter-tacking field, we have $(M/\beta)^2\ell(\ell+1)C_{\ell}^{\alpha\alpha}/2\pi \ll 10^{-6} {\rm deg}^2$. Note that the constraint of the parameter $(\beta/M)^2\lesssim O(10^{-3})$, which has been mentioned above. Considering this constraint, we get that $\ell(\ell+1)C_{\ell}^{\alpha\alpha}\ll 10^{-20} {\rm deg}^2$ for the slow-rolling case, and $\ell(\ell+1)C_{\ell}^{\alpha\alpha}\ll 10^{-10} {\rm deg}^2$ for the dark-matter-tacking case. These are all too small to be detectable, which are consistent with the results in previous works \cite{Caldwell:2011,lee}.

\section{Detection of the rotation angle fluctuations \label{sec4}}

The CMB polarization pattern can be decomposed into a gradient
E-mode and a curl B-mode component. In general, we can expand it
in terms of spin-weighted harmonics function as follows,
 \begin{equation}
 (Q\pm iU)(\hat{\bf n}) = \sum_{\ell m} a_{\pm 2,\ell m} ~_{\pm 2}Y_{\ell m}(\hat{\bf n}).
 \end{equation}
Then, the coefficients for the E-mode and B-mode polarizations can be constructed by
 \begin{equation}
 E_{\ell m}=-\frac{1}{2}(a_{2,\ell m}+a_{-2,\ell m}),~~B_{\ell m}=-\frac{1}{2i}(a_{2,\ell m}-a_{-2,\ell m}).
 \end{equation}
In the absence of the cosmological birefringence and cosmic weak
lensing, these two kinds of modes are independent, and satisfy the
Gaussian distribution. In particular, the B-mode can only be
generated by the relic gravitational waves \cite{ref-Bmode}, which
is proved to be much smaller than the E-mode component. In order
to simplify the problem, in this section, we assume the primordial
B-mode is zero in the last scattering surface (LSS). When the CMB
photons propagate from the LSS to us, due to the Chern-Simons term
in Eq.(\ref{csL}), the polarization plane of CMB is rotated, which
induces the mixture the these two modes. The rotation-induced
E-mode and B-mode are given by
\cite{li2008,kamionkowski2008,kamionkowski2009}
 \begin{equation}\label{delta-EB}
 \delta E_{\ell m}=2i\sum_{\ell_2 m_2} \sum_{\ell_3 m_3} \alpha_{\ell_2 m_2} E_{\ell_3 m_3} \xi_{\ell m \ell_3 m_3}^{\ell_2 m_2} H_{\ell \ell_3}^{\ell_2},
  ~~\delta B_{\ell m}=2\sum_{\ell_2 m_2} \sum_{\ell_3 m_3} \alpha_{\ell_2 m_2} E_{\ell_3 m_3} \xi_{\ell m \ell_3 m_3}^{\ell_2 m_2} H_{\ell \ell_3}^{\ell_2},
 \end{equation}
where, the functions $\xi_{\ell m \ell_3 m_3}^{\ell_2 m_2}$ and $H_{\ell \ell_3}^{\ell_2}$ are defined by
 \begin{equation}
 \xi_{\ell m \ell_3 m_3}^{\ell_2 m_2} \equiv (-1)^m\sqrt{\frac{(2\ell+1)(2\ell_2+1)(2\ell_3+1)}{4\pi}} {\left(\begin{array}{lcr}
\ell & \ell_2 & \ell_3 \\
-m & m_2 & m_3
\end{array}\right)},~~H_{\ell \ell_3}^{\ell_2}\equiv {\left(\begin{array}{lcr}
\ell & \ell_2 & \ell_3 \\
2 & 0 & -2
\end{array}\right)}.
 \end{equation}
Note that, for $\delta E_{\ell m}$, the only nonzero terms in the sum are those that satisfy $\ell+\ell_2+\ell_3=$even, while for $\delta B_{\ell m}$, the nonzero terms are those that satisfy $\ell+\ell_2+\ell_3=$odd. In principle, from Eq.(\ref{delta-EB}) and considering all the observation channels $EB$, $TB$, $EE$, $TE$, one can construct the best unbiased estimator for the rotation angle coefficients $\hat{\alpha}_{\ell m}$ as follows \cite{kamionkowski2008,kamionkowski2009,zaldarriaga2009},
 \begin{equation}\label{hat-alpha}
 \hat{\alpha}_{\ell m}= N_{\ell}^{\alpha\alpha}   \sum_{\ell_3\ge \ell_2} \sum_{AA'} G^{\ell}_{\ell_2 \ell_3}F_{\ell_2\ell_3}^{\ell,A'} \hat{D}_{\ell_2 \ell_3}^{\ell m,{\rm map}}[(\mathcal{C}^{\ell_2 \ell_3})^{-1}]_{AA'},
 \end{equation}
where the noise power spectrum is given by
 \begin{equation}\label{nl-alpha}
 N_{\ell}^{\alpha\alpha}=\left[\sum_{\ell_3\ge \ell_2} \sum_{AA'} G_{\ell_2 \ell_3}^{\ell} F^{\ell,A}_{\ell_2 \ell_3}F^{\ell,A'}_{\ell_2 \ell_3} [(\mathcal{C}^{\ell_2 \ell_3})^{-1}]_{AA'}\right]^{-1}.
 \end{equation}
In these expressions, the $AA'$ sums are over $\{ EB, BE, TB, BT, EE, TE, ET\}$ for $\ell_2 \neq \ell_3$ and $\{EB, TB, EE, TE\}$ for $\ell_2 =\ell_3$. The corresponding functions are defined as:
 $G_{\ell_2\ell_3}^{\ell}\equiv {(2\ell_2+1)(2\ell_3+1)}/{4\pi}$,
 $\hat{D}_{\ell_2 \ell_3}^{\ell m,XX',{\rm map}}=(G_{\ell_2\ell_3}^{\ell})^{-1} \sum_{m_2 m_3} X_{\ell_2 m_2}^{\rm map} {{X'}_{\ell_3 m_3}^{\rm map,}}^* \xi_{\ell_2 m_2\ell_3 m_3}^{\ell m}$, $\mathcal{C}_{AA'}^{\ell_2 \ell_3}\equiv G^{\ell}_{\ell_2 \ell_3}(\langle \hat{D}_{\ell_2 \ell_3}^{\ell m,A,{\rm map}}\hat{D}_{\ell_2 \ell_3}^{\ell m,A',{\rm map}}\rangle-\langle \hat{D}_{\ell_2 \ell_3}^{\ell m,A,{\rm map}}\rangle\langle\hat{D}_{\ell_2 \ell_3}^{\ell m,A',{\rm map}}\rangle)$ and $F_{\ell_2 \ell_3}^{\ell,A} \equiv 2Z^{A}_{\ell_2 \ell_3} H_{\ell_2 \ell_3}^{\ell} W_{\ell_2} W_{\ell_3}$. Note that, in these formulae, the spectrum $C_{\ell}^{XX',{\rm map}}$ is given by $C_{\ell}^{XX',{\rm map}} \equiv C_{\ell}^{XX'}W_{\ell}^2+N_{\ell}^{XX',{\rm noise}}$,
where $C_{\ell}^{XX'}$ is the rotated CMB power spectrum, $W_{\ell}$ is the beam window function of the CMB experiment, and $N_{\ell}^{XX',{\rm noise}}$ is the instrumental noise spectrum.

In reality, the CMB field at the LSS is an isotropic field, which
satisfies the Gaussian distribution. For a given multipole number
$\ell$, the $m$ components are statistically independent. So the
best unbiased estimator for the angle rotation power spectrum
$C_{\ell}^{\alpha\alpha}$ is
$\hat{C}_{\ell}^{\alpha\alpha}=\frac{1}{2\ell+1}\sum_{m}|\hat{\alpha}_{\ell
m}|^2$, and the corresponding variance is approximated by
 \begin{equation}
 \Delta \hat{C}_{\ell}^{\alpha\alpha}=\sqrt{\frac{2}{(2\ell+1)f_{\rm sky}}}(C_{\ell}^{\alpha\alpha}+N_{\ell}^{\alpha\alpha}),
 \end{equation}
where the term $C_{\ell}^{\alpha\alpha}$ stands for the cosmic
variance, which is quite important for the high-level detections,
but has been ignored in the work \cite{Caldwell:2011}. Note that,
the noise power spectrum $N_{\ell}^{\alpha\alpha}$ also depends on
$C_{\ell}^{\alpha\alpha}$ through its definition in
Eq.(\ref{nl-alpha}). Same with our previous work \cite{zhao2014},
in this paper, we consider the detection abilities of the Planck
satellite, CMBPol mission, and a `reference' experiment. For the
Planck satellite, we consider the best frequency at 143GHz
\cite{planck-bluebook}. For this channel, the beam Full width at
half maximum (FWHM) is $\theta_{\rm FWHM}=7.1'$. By 28-month
observation, the instrumental noises are expected to be
$N_{\ell}^{BB}=N_{\ell}^{EE}=2N_{\ell}^{TT}=2.79\times
10^{-4}\mu$K$^{2}$ \cite{planck-bluebook,ma2010}, and sky-cut
factor is adopted as $f_{\rm sky}=0.65$. For the potential CMBPol
mission, we consider the channel at 150GHz \cite{cmbpol}, where
beam FWHM is $5'$, the instrumental noises are
$N_{\ell}^{BB}=N_{\ell}^{EE}=2N_{\ell}^{TT}=0.83\times
10^{-6}\mu$K$^{2}$ \cite{cmbpol,zhao2011}, and sky-cut factor is
$f_{\rm sky}=0.8$. In this paper, a reference experiment is
considered as the far-future detector. We assume the detector
noise $\Delta_P=\sqrt{2}\Delta_T=1\mu{\rm K}$-${\rm arcmin}$
(corresponding the noise power spectrum
$N_{\ell}^{BB}=N_{\ell}^{EE}=2N_{\ell}^{TT}=0.85\times
10^{-7}\mu$K$^{2}$), beam FWHM $\theta_{\rm FWHM}=1'$ and the
sky-cut factor $f_{\rm sky}=1$. In Fig.\ref{figdd1}, we plot the
rotation noise power spectra $N_{\ell}^{\alpha\alpha}$ with red
dash-dotted lines. This figure shows that the detection of the
angle power spectrum mainly depends on the lower multipoles. Note
that, for the demonstration, in this figure we have assumed the
parameter $\beta H_I/M=0$ in the fiducial model.

\begin{figure}
\begin{center}
\includegraphics[width=10cm]{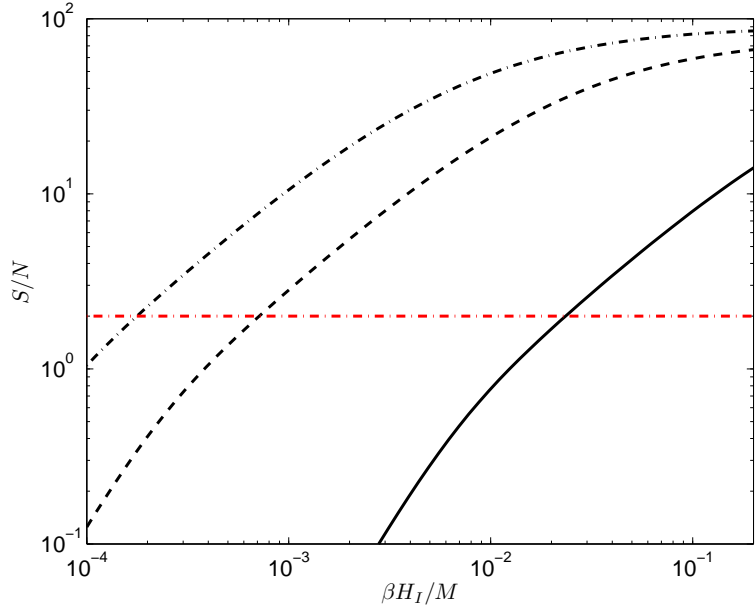}
\caption{\label{figbbb6} The signal-to-noise ratio $S/N$ as a function of the parameter $\beta H_I/M$. The solid line is for the Planck noise case, the dashed line is for the CMBPol noise case, and the dash-dotted one is for the case of the reference experiment.}
\end{center}
\end{figure}

In order to quantify the detection ability, as in general, we define the signal-to-noise ratio as follows,
 \begin{equation}
 S/N=\sqrt{\sum_{\ell}\left(\frac{C_{\ell}^{\alpha\alpha}}{\Delta \hat{C}_{\ell}^{\alpha\alpha}}\right)^2}.
 \end{equation}
In Fig. \ref{figbbb6}, we present the $S/N$ for all three cases. It shows that, if $\beta H_I/M=0.2$, which is the current upper limit value, Planck can definitely detect it, and the expected signal-to-noise ratio is $S/N=14$. In addition, we find that as long as $\beta H_I/M>0.024$, Planck will detect it at more than 2$\sigma$ level. For the potential CMBPol mission and the reference experiment, the detection abilities are much stronger. We find that, to get $S/N>2$, we only need $\beta H_I/M>7.2\times 10^{-4}$ for CMBPol, and $\beta H_I/M>1.8\times 10^{-4}$ for the reference experiment. So, we conclude that the future CMB observations will provide the excellent opportunity to detect the anisotropy of the cosmological birefringence.

\section{Rotated B-mode polarization and the de-rotating \label{sec5}}

\begin{figure}
\begin{center}
\includegraphics[width=10cm]{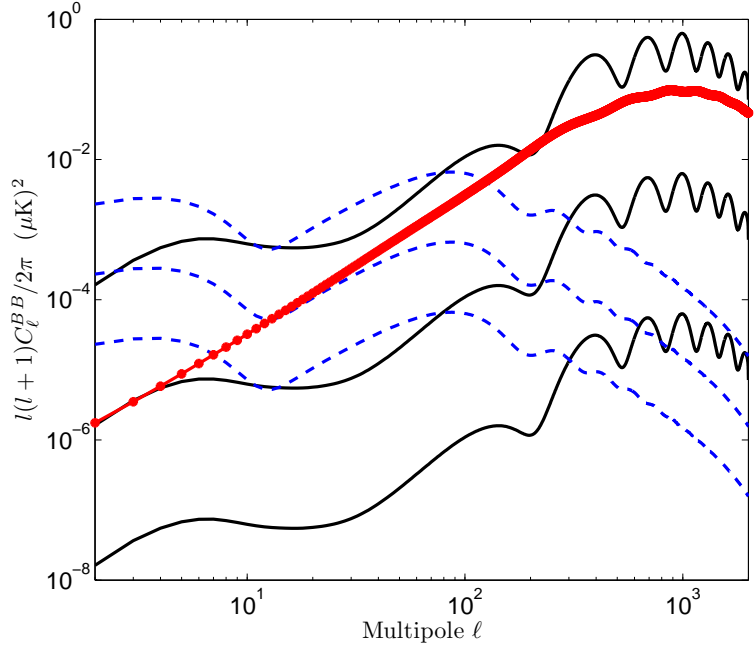}
\caption{\label{figbb3} The black solid lines show the rotated B-mode power spectrum. From the upper one to the lower one, we have adopted $\beta H_I/M=0.2$, $0.02$, $0.002$, respectively. The blue dashed lines denote the B-mode power spectra generated by primordial gravitational waves with $r=0.1$ (upper), $0.01$ (middle), $r=0.001$ (lower). The red curve denotes the B-mode power spectrum generated by cosmic weak lensing.}
\end{center}
\end{figure}

{\tc{As the main task of this paper, in this section, we shall focus on the B-mode polarization generated by the cosmological birefringence.}} To simply the problem, same to Sec. \ref{sec4}, we assume that the primordial (unrotated) B-mode polarization at the LSS is zero. By using the second formula in Eq.(\ref{delta-EB}), we get the rotated B-mode power spectrum \cite{li2008}
 \begin{equation}
 C_{\ell}^{BB}=\frac{1}{\pi}\sum_{\ell_2 \ell_3}(2\ell_2+1)(2\ell_3+1)C_{\ell_2}^{\alpha\alpha}C_{\ell_3}^{EE}(H^{\ell_2}_{\ell \ell_3})^2.
 \end{equation}
In Fig.\ref{figbb3}, we plot the spectrum $C_{\ell}^{BB}$ for the cases with $\beta H_I/M=0.2$, $0.02$, $0.002$, respectively. In addition, in this figure, the B-mode power spectrum generated by relic gravitational waves and the cosmic weak lensing are also plotted for the comparison. We find that, for the optimal case with $\beta H_I/M=0.2$, in all the multipole ranges, the rotated BB spectrum is compared, or even larger than, those caused relic gravitational waves and cosmic weak lensing, which definitely contaminates the gravitational-wave detection in the CMB. So, it is important to remove them in the future observations.

\begin{figure}
\begin{center}
\includegraphics[width=10cm]{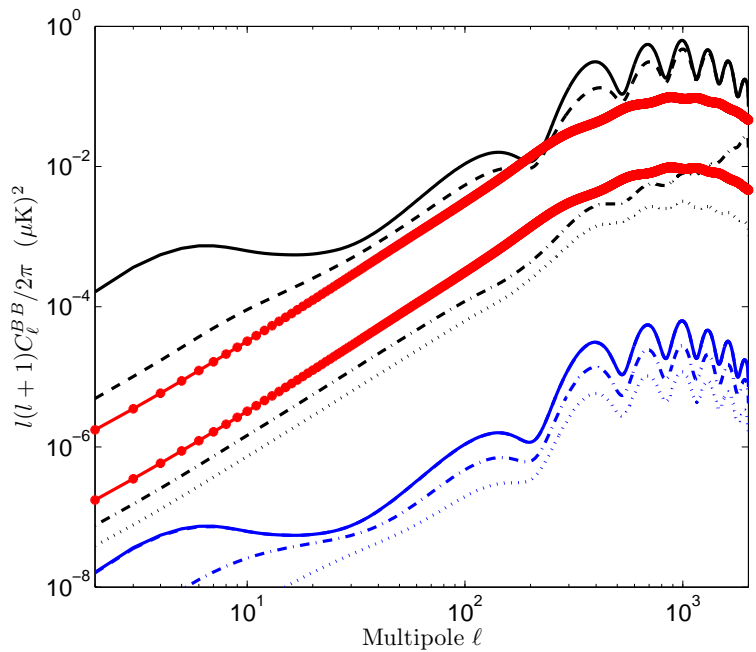}
\caption{\label{figbb4} The residual B-mode power spectrum after
the de-rotating. The black lines denote the results for the case
with $\beta H_I/M=0.2$, and the blue lines denote those for the
case with $\beta H_I/M=0.002$. In each case, the solid line shows
the spectrum without de-rotating, the dashed line shows the
residual spectrum after rotating, in which Planck noises are
considered, the dash-dotted line shows that for the CMBPol noise
case, and the dotted line shows that for the case of the reference
experiment. The red curve shows the original (upper line) and the
de-lensed (lower line) BB power spectra caused by the cosmic weak
lensing.}
\end{center}
\end{figure}

In the discussion in Sec. \ref{sec4}, we know that the rotation angle fluctuations can be recovered by the CMB observations. By using these reconstructions, similar to the de-lensing of CMB B-mode caused by the cosmic weak shear in \cite{knox2002,kamionkowski2002}, we can de-rotating the the rotated B-mode power spectrum as well. To preform de-rotating, we can construct the estimator $\hat{B}_{\ell m}$ for the rotated B-mode by utilizing the estimators $\hat{\alpha}_{\ell m}$ (defined in Eq.(\ref{hat-alpha})) and those of the noisy E-mode coefficients $\hat{E}_{\ell m}$, which can be written in the most general form as follows,
\begin{equation}
\hat{B}_{\ell m} = \sum_{\ell_2 m_2} \sum_{\ell_3 m_3} f(\ell,m,\ell_2,m_2,\ell_3,m_3) \hat{\alpha}_{\ell_2 m_2} \hat{E}_{\ell_3 m_3}.
\end{equation}
Note that the variances of $\hat{\alpha}_{\ell m}$ and $\hat{E}_{\ell m}$ are given by $\langle \hat{\alpha}_{\ell m} \hat{\alpha}^*_{\ell m}\rangle=C_{\ell}^{\alpha\alpha}+N_{\ell}^{\alpha\alpha}$, $\langle \hat{E}_{\ell m} \hat{E}^*_{\ell m}\rangle=C_{\ell}^{EE}+N_{\ell}^{EE}W_{\ell}^{-2}$. The coefficients $f$ are some general functions for $\ell,m,\ell_2,m_2,\ell_3$ and $m_3$. Thus, the residual power spectrum of the B-mode polarization is $C_{\ell}^{BB}({\rm residual})=\langle (B_{\ell m}-\hat{B}_{\ell m})(B_{\ell m}^{*}-\hat{B}_{\ell m}^{*})\rangle$. We minimize the residual power spectrum, and the coefficients $f$ are determined as follows,
 \begin{equation}
 f(\ell,m,\ell_2,m_2,\ell_3,m_3)= 2\xi^{\ell_2 m_2}_{\ell m \ell_3 m_3} H^{\ell_2}_{\ell \ell_3}{\Theta}_{\ell_2 \ell_3},
 \end{equation}
 where
 \begin{equation}
 {\Theta}_{\ell_2 \ell_3} \equiv \frac{C_{\ell_2}^{\alpha\alpha}}{C_{\ell_2}^{\alpha\alpha}+N_{\ell_2}^{\alpha\alpha}}
 \frac{C_{\ell_3}^{EE}}{C_{\ell_3}^{EE}+N_{\ell_3}^{EE}W_{\ell}^{-2}}.
 \end{equation}
The corresponding residual B-mode power spectrum is
 \begin{equation}
 C_{\ell}^{BB}({\rm residual})=\frac{1}{\pi}\sum_{\ell_2 \ell_3}(2\ell_2+1)(2\ell_3+1)C_{\ell_2}^{\alpha\alpha}C_{\ell_3}^{EE}(H^{\ell_2}_{\ell \ell_3})^2
 \left(1-{\Theta}_{\ell_2 \ell_3}\right).
 \end{equation}
As one can anticipate, for the ideal case, in which both noise
terms $N_{\ell}^{\alpha\alpha}$ and $N_{\ell}^{EE}$ are absent, we
have $\Theta_{\ell_2\ell_3}=1$, and $C_{\ell}^{BB}({\rm
residual})=0$, i.e., the rotated B-mode can be completely
reconstructed. In Fig.\ref{figbb4}, we plot the residual BB power
spectrum for the models with $\beta H_I/M=0.2$ (the upper limit
case) and $\beta H_I/M=0.002$. In the calculations, we have
considered the cases with Planck noises (dashed lines), CMBPol
noises (dash-dotted lines) and the reference experiment noises
(dotted lines), respectively. For the Planck noise case, the
de-rotating can only significantly reduce the B-mode spectrum at
the low multipoles if $\beta H_I/M$ is large. However, for the
reference experiment case, we find that the amplitudes of the
B-mode power spectrum in the all multipole ranges can be reduced
by more than two order for the model with $\beta H_I/M=0.2$, and
more than one order for the model with $\beta H_I/M=0.002$. It is
interesting to compare these residuals with BB power spectrum
caused by cosmic weak lensing. From Fig.\ref{figbb4}, we find
that, as long as the de-rotating can be done by considering the
noise level of CMBPol mission or a better experiment, the
residuals of cosmological birefringence are much smaller than
those of cosmic weak lensing.

The B-mode power spectrum caused by the cosmological birefringence
forms a new contamination for the detection of relic gravitational
waves in the CMB. Now, let us investigate whether or not the new
contamination can significantly influence the gravitational-wave
detection. To quantify it, we define the signal-to-noise ratio for
the gravitational waves \cite{zhao20092}
 $\left({S}/{N}\right)_{\rm gw} \equiv r/\Delta r$,
where $r$ is the tensor-to-scalar ratio, which characterizes the amplitude of relic gravitational waves, and $\Delta r$ is the observational uncertainty of $r$. For the detection with the CMB B-mode informational channel, the SNR can be calculated by \cite{zhao20092}
 \begin{equation}
 \left({S}/{N}\right)_{\rm gw}=\sqrt{\sum_{\ell}\left(\frac{C_{\ell,{\rm gw}}^{BB}}{\Delta \hat{C}_{\ell,{\rm gw}}^{BB}}\right)^2},
 \end{equation}
where $C_{\ell,{\rm gw}}^{BB}$ are the B-mode spectrum generated by the gravitational waves, and $\Delta \hat{C}_{\ell,{\rm gw}}^{BB}$ are the corresponding uncertainties of the estimators, which can be evaluated by $\Delta \hat{C}_{\ell,{\rm gw}}^{BB}=\sqrt{\frac{2}{(2\ell+1)f_{\rm sky}}}(C_{\ell,{\rm gw}}^{BB}+N_{\ell}^{BB})$. In order to focus on the influence of the cosmological birefringence, we assume $f_{\rm sky}=1$, and the noise power spectrum is $N_{\ell}^{BB}=C_{\ell}^{BB}({\rm residual})$, i.e., we only consider the residual B-mode caused by the the cosmological birefringence as the contaminations.

\begin{figure}
\begin{center}
\includegraphics[width=10cm]{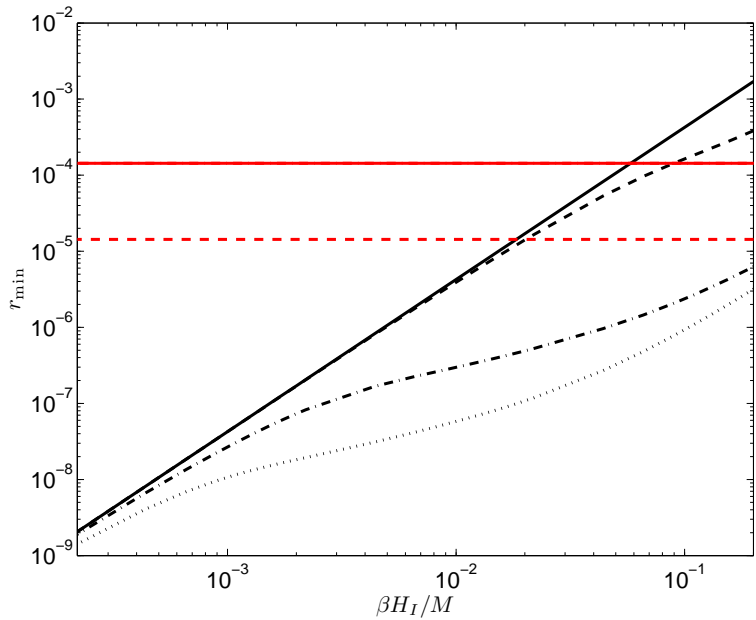}
\caption{\label{figbb1} The detection limits of the
tensor-to-scalar ratio $r$ if only considering the B-mode power
spectrum generated by cosmological birefringence as the
contaminations. The black solid line shows the results for the
case without de-rotating. The black dashed line shows that for
residual B-mode as the contaminations where the de-rotating is
performed by considering the Planck noise. The black dash-dotted
line and dotted line are exactly same with the dashed line, but
the CMBPol noises and reference experiment noises are considered,
respectively. For comparison, we have also plotted the detection
limits of $r$ if only considering the B-mode generated by cosmic
weak lensing (red solid line) and residual weak lensing (red
dashed line) as the contaminations.}
\end{center}
\end{figure}

Setting $(S/N)_{\rm gw}>2$, i.e., the gravitational waves can be detected at more 2$\sigma$ level, we solve the lower limit for the tensor-to-scalar ratio $r_{\rm min}$. In Fig.\ref{figbb1}, we plot the values of $r_{\rm min}$ for the models with different parameter values $\beta H_I/M$. We find that, without de-rotating, $r_{\rm min}=1.9\times 10^{-3}$ for the model with $\beta H_I/M=0.2$, which indicates that the gravitational-wave detection can be significantly limited by the cosmological birefringence, if the rotated B-mode polarization has not been properly reduced. However, if we de-rotate it by considering the CMBPol mission, the noises can be greatly reduced, and the limit of $r$ reduces to $r_{\rm min}=6.2\times 10^{-6}$, which corresponds the energy scale of inflation $V_{\rm min}=1.7\times 10^{15}$GeV. These limits could be even lower as $r_{\min}=3.1\times 10^{-6}$ and $V_{\rm min}=1.4\times 10^{15}$GeV, if the de-rotating is proceeded by considering the noise level of the reference experiment. Here, we should mention that, if only considering the B-mode caused by the residual cosmic weak lensing as the contamination, the detection limit of the gravitational waves is $r_{\rm min}=1.5\times 10^{-5}$ \cite{knox2002,kamionkowski2002,zhao20092}, which has been presented with red dashed line in Fig.\ref{figbb1}. So, consistent with the results in Fig.\ref{figbb4}, we conclude that, comparing with the residual cosmic weak lensing, the residual cosmological birefringence fluctuations cannot form a significant pollution for the detection of relic gravitational waves, which can be safely ignored in the future detections.

\section{Conclusions \label{sec6}}

The detection of relic gravitational waves through the CMB B-mode polarization is rightly considered a highest priority task for the future observational missions. In addition to the contaminations and noises in the observations, in the real Universe, the primordial B-mode was also polluted by some other cosmological effects. One of them is the cosmological birefringence, which can be caused by the possible coupling between CMB photons and the cosmic scalar field through the Chern-Simons term. The cosmological birefringence rotates the polarization planes of CMB photons, and mixes the E-type and B-type polarizations. Due to the perturbations of the scale field, the fluctuations of the rotation angle naturally exist.

In this paper, we studied the fluctuations of cosmological
birefringence in various scalar-field models. {\tc{By detailed discussing both entropy and adiabatic perturbation modes for the first time,
we find that, if the cosmic scalar field is identified with the quintessence component,
the fluctuations of the rotation angle are always too small, and beyond the detection
abilities of future detectors. However, if the scalar field is a
massless field, consistent with the previous works, we find that the fluctuation power spectrum could be fairly
large and detectable for the potential observations, although the
uniform rotation angle is absent in this case.}}

{\tc{As the main task of this paper, the B-mode polarization caused by the rotation angle fluctuations
are also studied in details. We found that, the converted B-mode
polarizations could be comparable with (or even larger than) those
generated by the primordial gravitational wave or those caused by
the cosmic weak lensing. However, this B-mode can be well
reconstructed and subtracted at the very high level. In this paper, we proposed the method to de-rotate the
B-mode polarization by utilizing the statistical properties of the coefficients $\alpha_{\ell m}$ and $E_{\ell m}$.
We found that, if the de-rotating is done by considering the noise level of
CMBPol mission or the better experiment, the B-mode residuals
become much smaller than those caused by weak lensing, which is
entirely negligible for the future gravitational-wave detections.}}

~

~

{\tc{Note: In the same day of the submission of the paper, BICEP2 \cite{bicep2} released its data which indicates a discovery of the primordial gravitational waves with $r = 0.20^{+0.07}_{-0.05}$ and $r = 0$ disfavored at $7.0\sigma$, which confirmed the hint of gravitational waves in the WMAP and Planck low-multipole data \cite{zhao2010,zhao2014}.}}

~

~

{\it Acknowledgments:}
W.Z. is supported by project 973 under Grant No.2012CB821804, by NSFC No.11173021, 11322324 and project of KIP of CAS. M.L. is supported by Program for New Century Excellent Talents in University and by NSFC under Grants No. 11075074.

\appendix

\section{Numerical calculation of the perturbation $\delta\varphi_k$}

The anisotropies of cosmological birefringence is determined by the synchronous-gauge scalar-field fluctuation $(\delta\varphi)_{\rm syn}$ at the decoupling stage \cite{li2008,Caldwell:2011}. To obtain this fluctuation, we start from the initial condition from inflation, and then evolve the scalar field perturbation equation of motion forward in time, from early radiation-dominated epoch to the decoupling time. In order to use the analytical approximations for the cosmological perturbations, which are shown in Appendix B, we shall start in the conformal Newtonian gauge, and then transfer the results to the synchronous gauge. In the conformal Newtonian gauge, the perturbations are characterized by two scalar potentials $\Psi$ and $\Phi$. The line element is written as
 \begin{equation}
 ds^2=a^2(\tau)\left\{ -(1+2\Psi)d\tau^2+(1-2\Phi)dx^i dx_i\right\}.
 \end{equation}
In this gauge, the equation of motion is,
 \begin{eqnarray}\label{appda-1}
 {\delta\varphi^{\rm n}_k}''+2\mathcal{H}{\delta\varphi^{\rm n}_k}'+a^2 V_{\bar{\varphi}\bar{\varphi}}{\delta\varphi^{\rm n}_k}+k^2{\delta\varphi^{\rm n}_k}=\bar{\varphi}'(3\Phi_k'+\Psi_k')-2a^2V_{\bar{\varphi}}\Psi_k.
 \end{eqnarray}
Note that, the perturbation ${\delta\varphi^{\rm n}_k}$ here denotes the quantity in the conformal Newtonian gauge, i.e., ${\delta\varphi^{\rm n}_k}\equiv (\delta\varphi_k)_{\rm newtonian}$, which relates to the corresponding quantity in the synchronous gauge by the following formula \cite{Ma:1995}
 \begin{eqnarray}\label{gauge}
 (\delta\varphi_k)_{\rm syn}={\delta\varphi^{\rm n}_k}-\gamma_k\bar{\varphi}',
 \end{eqnarray}
where $\gamma_k$ is determined by the equation
$\gamma_k'+\mathcal{H}\gamma_k=\Psi_k$ and
$\gamma_k=(1/2)\Psi_k/\mathcal{H}$ in the radiation domination and
$\gamma_k\simeq (2/3)\Psi_k/\mathcal{H}$ in the matter domination.
Now let us search the adiabatic mode. Before the exactly numerical
calculations, we first neglect the effects of anisotropic stress,
and study the general properties of the solution. In this case, we
have $\Phi_k=\Psi_k$, and Eq. (\ref{appda-1}) can be rewritten as
\cite{zhaogb} \bea\label{eom}
\zeta_{\varphi}'&=&-\frac{\theta_{\varphi}^{\rm n}}{3}-[\frac{w'}{1+w}+3\mathcal{H}(1-w)](\zeta_{\varphi}+\Psi_k+\frac{\mathcal{H}}{k^2}\theta^{\rm n}_{\varphi})~,\nonumber\\
\theta'^{\rm n}_{\varphi}&=&2\mathcal{H}\theta_{\varphi}+k^2(3\zeta_{\varphi}+4\Psi_k)~,
\eea
where $\zeta_{\varphi}=-\Psi_k-\mathcal{H}\delta\rho_{\varphi}/\rho_{\varphi}'$ is the curvature perturbation of the surface of homogeneous
quintessence density, and $\theta$ corresponds to the momentum perturbation (the definition see \cite{Ma:1995}), for the scalar field $\theta^{\rm n}_{\varphi}=k^2\delta\varphi_k^{\rm n}/\varphi'$. The curvature perturbation is gauge invariant, but $\theta$ is not.
The metric perturbations at large scales are determined by the dominant fluid,
\be
\Psi_k\simeq -\zeta_f-\frac{\mathcal{H}}{k^2}\theta_f^{\rm n}~,
\ee
where the subscript $f$ represents the dominant fluid, radiation or matter. The solution is
\bea
\zeta_f&=& -\frac{5+3w_f}{3(1+w_f)}\Psi_k=constant~,\nonumber\\
\theta_f^{\rm n} &=& \frac{2k^2}{3}\frac{\Psi_k}{\mathcal{H}(1+w_f)}~.
\eea
For constant $w_f$, the reduced Hubble parameter is $\mathcal{H}=2/[(1+3w_f)\tau]$. It is easy to see from these equations that there is a special solution to the equation (\ref{appda-1}) or equivalently the equations (\ref{eom}), which is at large scales $k\tau\ll1$ (please note that numerically $\theta^{\rm n}\sim \mathcal{O}(k^2)\zeta$) approximately:
\be
\zeta_{\varphi}=\zeta_f~,~\theta^{\rm n}_{\varphi}=\theta_f^{\rm n}~.
\ee
These are the usual adiabatic condition, so we called this special solution the adiabatic mode (we assume there is no dark matter entropy perturbation, hence $\zeta_m=\zeta_r$).
With these formula one can check that when converting to the synchronous gauge in terms of Eq. (\ref{gauge}),
$(\delta\varphi_{k})_{\rm syn}$ vanishes at both the radiation and matter dominated epoches.
In reality, some neglected quantities in above analysis such as the density of quintessence and the anisotropic stress do not vanish,
$(\delta\varphi_{k})_{\rm syn}$ in the adiabatic mode should be small but not zero. These are included in the numerical computations.

In the real calculation, we shall solve Eq.(\ref{appda-1}), where
the terms $a^2 V_{\bar{\varphi}}$ and $a^2
V_{\bar{\varphi}\bar{\varphi}}$ can be calculated by using the
relations in Eqs.(\ref{v-a}) and (\ref{v-aa}), respectively. The
background component of quintessence is govern by the motion of
equation in Eq.(\ref{varphi0aaa}). The initial condition of this
equation is determined by the current quintessence energy density
and the equation of state: $\Omega_{quint}=0.685$,
$w=w_0+w_a(1-a)$. The other cosmological parameters are adopted as
follows \cite{planck} $\Omega_c=0.266$, $\Omega_b=0.0487$,
$h=0.673$, $\tau_{reion}=0.089$, $n_s=0.9603$,
$\ln(10^{10}A_s)=3.089$. For the cosmological perturbations
$\Psi_k$ and $\Phi_k$, we adopt the analytical approximations (see
Appendix B), which are excellently consistent with the exact
numerical results \cite{hu}.

Thus, we can numerically solve Eq.(\ref{appda-1}). The initial conditions are set at the beginning of radiation-dominant stage, where ${\delta\varphi^{\rm n}_k}'=0$ and $P_{\delta\varphi^{\rm n}}(k,\tau_{ini})=H_I^2/(4\pi^2)$. For the quintessence models with given $w_0$ and $w_a$, we calculate $(\delta\varphi_k)_{\rm syn}$, and substitute into Eq.(\ref{clalpha}) to obtain the power spectrum $C_{\ell}^{\alpha\alpha}$. The results are presented in Fig.\ref{figbbbb1}.

\section{Analytical approximations of the cosmological perturbations}

In this section, we simply summarize the analytical approximation of the cosmological perturbations, which have been used in Appendix A. The details of these results can be found in the previous work \cite{hu}.

In both radiation-dominant and matter-dominant stages, the
perturbations in the conformal Newtonian gauge can be approximated
as follows,
 \begin{eqnarray}
 \Phi_k(a_{h})=\bar{\Phi}_k(a_{h})\left\{ [1-T_f(k)]\exp[-\alpha_1(a_{h} k/ k_{eq})^{\beta}] +T_f(k)\right\},
 \end{eqnarray}
 \begin{eqnarray}
 \Psi_k(a_{h})=\bar{\Psi}_k(a_{h})\left\{ [1-T_f(k)]\exp[-\alpha_2(a_{h} k/ k_{eq})^{\beta}] +T_f(k)\right\},
 \end{eqnarray}
where $\alpha_1=0.11$, $\alpha_2=0.097$, $\beta=1.6$,
$k_{eq}=(2\Omega_0 H_0^2 a_0)^{1/2}$. $a_h$ is the scale factor,
which has been normalized at the matter-radiation equality time.
The transfer function is given by
 \begin{eqnarray}
 T_f(k)=\frac{\ln (1+2.34q)}{2.34q}\left[1+3.89q+(14.1q)^2+(5.46q)^3+(6.71q)^4\right]^{-1/4},
 \end{eqnarray}
where $q\equiv \frac{k}{\Omega_0 h^2 e^{-2\Omega_b}}$. The initial
conditions are given by
 \begin{eqnarray}
 \bar{\Phi}_k(a_{h})=\frac{3}{4}\left(\frac{k_{eq}}{k}\right)^2 \frac{a_{h}+1}{a_{h}^2}\left[1+\frac{2}{5}f_{\nu}(1-0.333\frac{a_{h}}{a_{h}+1})\right]A U_{G}(a_{h}),
 \end{eqnarray}
and
 \begin{eqnarray}
 \bar{\Psi}_k(a_{h})=-\frac{3}{4}\left(\frac{k_{eq}}{k}\right)^2 \frac{a_{h}+1}{a_{h}^2}
 \left\{\left[1+\frac{2}{5}f_{\nu}(1-0.333\frac{a_{h}}{a_{h}+1})\right]A U_{G}(a_{h})
 +\frac{8}{5}f_{\nu}\frac{\bar{N}_2(a_{h})\cos(0.5k/Ha)}{a_{h}+1}\right\},
 \end{eqnarray}
where the Hubble parameter is $H(a_{h})=\frac{{a}'_{h}}{a_{h}} \frac{a_{h 0}}{a_{h}}$, and the fraction of neutrino is $f_{\nu}=0.405$. The functions $U_G$ and $\bar{N}_2$ are given by
 \begin{eqnarray}
 U_{G}(a_h)=\left[a_h^3+\frac{2}{9}a_h^2-\frac{8}{9}a_h-\frac{16}{9}+\frac{16}{9}\sqrt{a_h+1}\right]\frac{1}{a_h(a_h+1)},
 \end{eqnarray}
 \begin{eqnarray}
 \bar{N}_2(a_h)=-\frac{1}{10}\frac{20 a_h+19}{3a_h+4} A U_G-\frac{8}{3}\frac{a_h}{3a_h+4} A +\frac{8}{9}\ln \left(\frac{3a_h+4}{4}\right)A.
 \end{eqnarray}
The parameter $A$ is determined by the power spectrum of primordial density perturbations, which is
 \begin{eqnarray}
 P_{s}(k)=\frac{k^3 \Phi_k^2(\tau_{ini})}{2\pi^2} \left(\frac{1.5+0.4 f_{\nu}}{1+0.4 f_{\nu}}\right)^2 =A_s (k/k_0)^{n_s-1},
 \end{eqnarray}
where $\Phi_k(\tau_{ini})=\frac{5}{6}(1+0.4 f_{\nu})(k_{eq}/k)^2 A$.

\end{document}